# Non-linear lattice response of SmFeAsO$_{1-x}$F$_x$ superconductor to hydrostatic pressure

E. Liarokapis[1]*, M. Calamiotou[2], N. D. Zhigadlo[3], S. Katrych[3], J. Karpinski[3]

[1]*Department of Physics, National Technical University of Athens, GR15780, Athens, Greece*
[2]*Department of Physics, University of Athens, Athens, Greece*
[3]Laboratory for Solid State Physics, ETH Zurich, 8093 Zurich, Switzerland

ABSTRACT

Hydrostatic pressure Raman measurements have been carried out on the SmFeAsO series of oxypnictides with varying amount of doping (F substitution for O and Co for Fe) and transition temperature ($T_c$), in order to investigate lattice modifications and their connection to doping and superconductivity. Synchrotron XRD data on some of these compounds indicated that at low doping the lattice constants vary smoothly with pressure, but with increasing F concentration there is a deviation from the normal equation of state and these effects are related with modifications in the superconducting FeAs$_4$ tetrahedra. The hydrostatic pressure Raman measurements show that the A$_{1g}$ mode of the rare earth atom for the superconducting compounds deviates from the linear pressure dependence at the same pressures where the XRD results indicate pressure-induced lattice anomalies. A similar anomaly is found for the As phonon of same symmetry. As in cuprates, the effect is diminished in the undoped compounds and it is not related with the F substitution being present in the Sm(Fe$_{1-x}$Co$_x$)AsO as well. The calculated Grüneisen parameter for the Sm phonon ($\gamma \approx 1.5$) is very similar to the corresponding values of cuprates and it does not vary with doping. For the Fe mode it has higher value ($\gamma \approx 1.8$) than for As ($\gamma \approx 1$) indicating a more anharmonic phonon.

___

*Corresponding author email: eliaro@central.ntua.gr

# 1. INTRODUCTION

The effect of hydrostatic pressure on the transition temperature ($T_c$) of the high temperature superconductors has been used extensively [1], as it can indicate possible atomic substitutions for the preparation of compounds with higher $T_c$. Besides, it can help the study of the superconductivity property and its relation to lattice by modifying the interatomic distances and driving the high $T_c$ superconductors towards or away from their lattice instabilities [2, 3]. While there is an accumulated amount of data from these studies [1, 4], the investigation of high-pressure effects on the Raman spectra is limited. The main reason is the difficulty of studying the very weak modes of the superconducting materials inside the high-pressure cell.

The discovery of high $T_c$ superconductivity in the Fe pnictides has triggered a large amount of research since these materials have a similar phase diagram with the cuprates [5]. The compounds have a structural phase transition and a spin density wave at low temperatures that varies depending on the amount of doping and the rare earth [5]. Indications for a lattice instability at low temperatures has been found in oxygen deficient $NdFeAsO_{0.85}$ compound with the maximum $T_c$ originally from a FTIR study [6] that has been confirmed by low temperature synchrotron measurements on the same compounds [7] and in F substituted ones [8]. The Raman measurements on the same samples were not conclusive and only indicated a small softening of the As mode at low temperatures [6]. Up to now, the results of Raman measurements in pnictides are controversial [9-11], [12], which may be due to the sensitivity of the pnictides to local heating that hide the small effects in the Raman spectra.

Based on the similarity of the pnictides with the cuprates and the fact that lattice instabilities in the cuprates are correlated with modifications in $T_c$ [2,3], we have decided to carry out a hydrostatic pressure study of the SmFeAsO series of pnictides with F substitution for O or Co for Fe. The effect of hydrostatic pressure on the transition temperature of these compounds resembles the situation in the high $T_c$ cuprates [1,4]. But contrary to the cuprates, there are indications that undoped SmFeAsO becomes superconducting at high pressures [13]. Apart from that, the general behavior of the pnictides follows the rule, that the $T_c$ of the underdoped compounds increases with pressure and decreases in the overdoped ones [11].

## 2. EXPERIMENTAL DETAILS

Single crystals of non-superconducting SmFeAsO and superconducting $SmFe_{0.92}Co_{0.08}AsO$ ($T_c$=16.4K) and polycrystalline $SmFeAsO_{1-x}F_x$ ones of various doping levels have been studied. The preparation of the samples is described elsewhere [14-16]. Typical size of the single crystals was a few hundred micrometers while for the polycrystals of the order of a few microns. For the Raman measurements a Jobin-Yvon triple spectrometer equipped with a microscope (×40 magnification) and a liquid $N_2$ cooled charge coupled device (CCD) and the 514.5 nm line from an $Ar^+$ laser were used. Typical accumulation times were 1-2 hours depending on the scattering polarization and the material. The local heating due to the laser beam is expected to be small due to the low laser power used (less than 0.1mW) and the immediate contact of the sample with the pressure-transmitting medium. The pressure and information about a non-hydrostatic environment was obtained from a small silicon crystal inside the DAC. Synchrotron angle-dispersive powder diffraction patterns at high pressure have been collected at the beamline XRD1 of Elettra, SincrotroneTrieste, Italy. Merrill-Bassett type diamond anvil cells (DAC) were used with a mixture 4:1 of methanol-ethanol as pressure transmitting medium, which for such low pressures is expected to be hydrostatic. More details about the synchrotron measurements have been published elsewhere [17].

## 3. DATA ANALYSIS AND DISCUSSION

In the P4/nmm structure the Raman active phonons of the 1111 pnictides consist of $2A_{1g}$, $2B_{1g}$, and $4E_g$ symmetry modes from which the first four have eigenvectors along the c-axis, while the other four along the ab-plane. The strongest modes are the $A_{1g}$ that involve vibrations of the rare earth or the As atoms, and the $B_{1g}$ of the Fe or O atoms [18]. All modes are much weaker than the relative modes of cuprates and sensitive to laser power.

Figures 1 presents the hydrostatic pressure dependence of the frequency of the non-superconducting undoped SmFeAsO for the phonons due to Sm, As, and Fe. Figure 2 shows the characteristic dependence of the width for the Sm mode. For this phonon the blue circles indicate the data with increasing pressure and the red triangles for decreasing. It is clear that there is no hysteresis and this holds for all data presented below. Up to the highest pressure studied (6 GPa) the frequency of all modes increases smoothly with pressure and at least for As and Fe there is no apparent deviation from linearity within statistics. For the stronger Fe mode, for which the better statistics provided more accurate results, there is a small contribution from a second order term to pressur, in close agreement with the XRD results [17]. Since we could not reach higher pressures, it is unclear whether there will be any lattice anomaly at higher pressures, where the compound may become superconducting [13]. The spread of data for the bandwidth of Sm and for the other two modes (not shown in Fig.2) cannot provide any clue for irregular behavior.

Our structural data on the SmFeAsO$_{1-x}$F$_x$ compounds have been published elsewhere [17] and agree with other high pressure XRD measurements where lattice modifications were detected correlated with changes in T$_c$ [19]. Figure 3 presents the position of As atom from these data together with the high pressure Raman data of the nearly optimally doped SmFeAsO$_{0.83}$F$_{0.17}$ (T$_c$=44K). Only the pressure dependence of the Sm and As modes is shown in Fig.3 since the Fe mode was not strong enough to provide reliable data. Both the As and the Sm phonons show an increase in frequency up to ~1-1.4 GPa and above ~2.5-3 GPa, while in the intermediate pressure ~1.2-2.7 GPa the frequency remains roughly constant. A similar anomalous behavior is observed in the position of the As atom, which indicates that a lattice modification is induced at the ~1.2-2.5 GPa pressure range within the superconducting FeAs$_4$ tetrahedra. Our data on the oxypnictides are reminiscent of the anomalous non-linear behavior observed in the hydrostatic pressure Raman studies of the YBCO cuprates [20], which had been verified by synchrotron XRD measurements [2,3].

Although the results of the Rietveld analysis presented in Fig.3 have considerable spread, an analogous lattice modification can be deduced from the lattice parameter presented in Figure 4. In this graph the relative variation of c- and a-axis with pressure is shown for both the undoped and the superconducting compound (x=0.17). While in the undoped SmFeAsO there is no appreciable deviation and the two axis are compressed isotropically, the doping induces an internal anisotropic behavior in the pressure range ~1.2 to ~2.7 GPa. This is the same pressure limits where the Fe and As phonons show the anomaly (Fig.3) indicating that the reason of this effect is a doping dependent anisotropic response of the two axis to pressure and more specifically a non-monotonic compression of the superconducting $FeAs_4$ layer [17], apparently modifying the distribution of the carriers. In addition, our results provide a further indication for the connection of this effect to doping with the data pointing to a local lattice instability very similar to those observed before by high pressure studies in the cuprates [2,3,20]. Although more data are necessary to verify the tendency to local structural modifications and their dependence on doping, the absence of a similar effect in the non-superconducting SmFeAsO (Fig.1) and $PrBa_2Cu_3O_7$ [3] compounds are strong indications that the lattice distortions are directly related with doping.

As an independent test of the lattice instability and the possibility to be induced by the lattice defects from the substitution, we have studied the superconducting single crystal of $SmFe_{0.92}Co_{0.08}AsO$. Figures 5 and 6 present the frequency and width of the Sm phonon under pressure. It is clear that there is a very similar situation with the superconducting compound $SmFeAsO_{0.83}F_{0.17}$ and the frequency increases up to roughly 2.5 GPa, to be followed again by a plateau up to ~3.5 GPa and then to increase again. A strong anomaly is also observed in the width of the phonon, which decreases with pressure with a sudden increase at ~2.5 GPa to be followed by another drop in width (Fig.6). This is roughly the same pressure where the phonon frequency shows a pressure-induced anomaly (Fig.5). Again the behavior of both frequency and width are indicative of lattice instability.

Concerning the softness of the bonds, the slope of the frequency for the Sm atom (dln$\omega$/dp) is the same (0.015) for all compounds studied, and therefore it can be considered as independent of doping. The slope of the Fe mode has been measured with accuracy only for the undoped compound and it has a larger value (0.019), while for the As atom it varies from 0.011 (for the undoped) to 0.009 (for x=0.17). It should be noted that for the x=0.17 compound the slopes were calculated as a linear fit to all data, ignoring the trend of the data to three different regions as depicted in Fig.3. The smaller value of the slopes for the As atom indicate stiffer bonds, which agrees with the results of EXAFS [21]. Based on the measured value of the bulk modulus measured in Ref.[17], we could calculate the Grüneisen parameter for the three modes, which are given in Table I and show that for Sm it is $\gamma$=1.45±0.05 and independent of doping, for As is quite smaller ($\gamma$=1.00±0.09 for x=0.0 to 0.85±0.08 for x=0.17) and may depend slightly on doping, while for Fe (measured only for SmFeAsO) seems to be larger ($\gamma$=1.85±0.2). The low value of the Grüneisen parameter for the As phonon and its possible reduction with doping agrees with the observation that the As bond length get harder in the superconducting regime [21]. The larger value of the Sm mode is indicative of the more ionic character of the SmO layer while its value agrees with the values observed in many metal oxides [22]. Finally, the even larger value of the Fe phonon should be due to an increased anharmonicity of this mode compared to Sm and As, as the low temperature Raman measurements also indicate (stronger softening of this mode with temperature, Ref.[23]). The considerable difference between the Grüneisen parameters must be related with the observed anisotropic compression of the FeAs$_4$ tetrahedra [17].

Similar values for the Grüneisen parameters have been obtained for the YBa$_2$Cu$_3$O$_x$ cuprates (Ref.[24], which agrees with the average pressure dependence of our data presented in Ref.[20]). Although the original high pressure study of Ref.[24] has not revealed any phonon anomalies, our more recent studies had clearly shown that the strong A$_g$ phonons follow a non-linear pressure dependence [20], similar to the one observed in this work. It therefore appears that in pnictides as in cuprates this instability is intrinsic, it is related with the free carriers being absent in the non-superconducting compound, while it is not connected with any lattice defects from atomic substitutions. As a general rule, in the high temperature superconductors the application of hydrostatic pressure increases $T_c$ towards optimal doping and even in the undoped compound of SmFeAsO pressure can induced superconductivity [13]. It is reasonable to assume that around optimal doping the lattice exhibits an instability and the amount of pressure required to reach such instability decreases with doping. In any case, a connection between lattice distortions and superconductivity seems to be present in this family of pnictides as in cuprates.

## 4. CONCLUSIONS

We have measured the pressure dependence of certain Raman active modes of undoped and superconducting iron pnictide 1111 compounds SmFeAsO with F substitution for O and Co for Fe in order to investigate possible lattice anomalies and their dependence on doping, as observed in the cuprates with similar phase diagram characteristics. We have found clear lattice modifications, which are absent in the undoped non-superconducting compound and do not depend on the F substitution for O. Our results fully agree with our synchrotron XRD measurements on the same compounds, which indicate an anisotropic compression of the unit cell and more specifically of the $FeAs_4$ tetrahedra that are responsible for superconductivity. Furthermore, we have measured the Grüneisen parameters of the stronger modes and found them to vary considerably between the As and the Fe phonons justifying the anisotropic compression of the unit cell. It appears that the lattice instability is inherent in this family of superconductors as in cuprates indicating that lattice distortions related to doping are intrinsic for both families of high $T_c$ superconductors.

Table I

| Grüneisen parameters | | | |
|---|---|---|---|
| Compound | Phonon mode | | |
| | Sm | As | Fe |
| SmFeAsO | 1.44±0.04 | 1.00±0.09 | 1.85±0.2 |
| SmFeAsO$_{0.83}$F$_{0.17}$ | 1.52±0.08 | 0.85±0.08 | |
| SmFe$_{0.92}$Co$_{0.08}$AsO | 1.43±0.05 | | |

**REFERENCES**

[1] J.S. Schilling, "Treatise on High Temperature Superconductivity", J.R. Schrieffer, editor (Springer Verlag, Hamburg, 2006).

[2] E. Liarokapis, D. Lampakis, E. Siranidi, and M. Calamiotou, J. of Phys. and Chem. of Solids **71**, 1084 (2010).

[3] M. Calamiotou, A. Gantis, E. Siranidi, D. Lampakis, J. Karpinski, E. Liarokapis, Phys. Rev. B **80**, 214517 (2009).

[4] C.W. Chu and B. Lorenz, Physica C **469**, 385 (2009).

[5] M.R. Norman, Physics 1, 21 (2008).

[6] E. Siranidi et al., Journal of Alloys and Compounds **487**, 430 (2009).

[7] M. Calamiotou, I. Margiolaki, A. Gantis, E. Siranidi, Z.A. Ren, Z.X. Zhao, and E. Liarokapis, Europ. Phys. Lett. 91, 57005 (2010).

[8] H. Fujishita, Y. Hayashi, M. Saito, H. Unno, H. Kaneko, H. Okamoto, M. Ohashi, Y. Kobayashi, and M. Sato, Eur. Phys. J. B **85**, 52 (2012).

[9] L. Chauvière, Y. Gallais, M. Cazayous, A. Sacuto, and M. A. Méasson, D. Colson and A. Forget, Phys. Rev. B**80**, 094504 (2009).

[10] M. Rahlenbeck, G. L. Sun, D. L. Sun, C. T. Lin, B. Keimer, and C. Ulrich, Phys. Rev. B**80**, 064509 (2009).

[11] P. Kumar, A. Bera, D. V. S. Muthu, A. Kumar, U. V. Waghmare, L. Harnagea, C. Hess, S. Wurmehl, S. Singh, B. Buechner, and A. K. Sood, J. Phys.: Condens. Matter **23**, 255403 (2011).

[12] Y. Gallais, A. Sacuto, M. Cazayous, P. Cheng, L. Fang, and H. H. Wen, Phys. Rev. B**78**, 132509 (2008).

[13] H. Takahashi, H. Okada, Y. Kamihara, S. Matsuishi, M. Hirano, H. Hosono, K. Matsubayashi, Y. Uwatoko, J. Phys.: Conf.Series **215**, 012037 (2010).

[14] N.D. Zhigadlo, S. Katrych, Z. Bukowski, S. Weyeneth, R. Puzniak, and J. Karpinski, J. Phys.: Condens. Matter **20**, 342202 (2008).

[15] J. Karpinski, N.D. Zhigadlo, S. Katrych, Z. Bukowski, P. Moll, S. Weyeneth, H. Keller, R. Puzniak, M. Tortello, D. Daghero, R. Gonnelli, I. Maggio-Aprile, Y. Fasano, O. Fischer, K. Rogacki, and B. Batlogg, Physica C **469**, 370 (2009).

[16] N.D. Zhigadlo, S. Weyeneth, S. Katrych, P.J.W. Moll, K. Rogacki, S. Bosma, R. Puzniak, J. Karpinski, and B. Batlogg, arXiv: 1208.6207.

[17] M. Calamiotou, D. Lampakis, E. Siranidi, J. Karpinski, N.D. Zhigadlo and E. Liarokapis, to appear in Physica C (2012).

[18] V. G. Hadjiev, M. N. Iliev, K. Sasmal, Y.-Y. Sun, and C.W. Chu, Phys. Rev. B**77**, 220505 (R) (2008).

[19] G. Garbarino, R. Weht, A. Sow, A. Sulpice, P. Toulemonde, M. Alvarez-Murga, P. Strobel, P. Bouvier, M. Mezouar, M. Nunez-Regueiro, Phys. Rev. B **84**, 024510 (2011).

[20] E. Liarokapis in "Symmetry and Heterogeneity in High Temperature Superconductors", Ed. A. Bianconi, p. 117-132 (Springer, 2006).

[21] B Joseph, A Iadecola, L Malavasi, and N L Saini, J. Phys.: Condens. Matter **23**, 265701 (2011).

[22] H. Ledbetter, Physica C**159**, 488 (1989).

[23] E. Liarokapis, A. Antonakos, N. D. Zhigadlo, S. Katrych, and J. Karpinski, Conference Proceedings Volume of ICSM-2012.

[24] K. Syassen, M. Hanfland, K. Strössner, M. Holtz, W. Kress, M. Cardona, U. Schröder, J. Prade, A.D. Kulkarni, and F.W. de Wette, Physica C**153-155**, 264 (1988).

**FIGURE CAPTIONS**

Figure 1. Hydrostatic pressure dependence of Sm, As, and Fe phonon frequencies of SmFeAsO. Straight lines are linear fits to the data. For the Sm mode a $2^{nd}$ order fit could produce better fitting, but for the other modes such fitting would be meaningless due to the spread of the data.

Figure 2. Hydrostatic pressure dependence of the Sm phonon width of SmFeAsO.

Figure 3. Hydrostatic pressure dependence of Sm and As phonon frequency and the As height of SmFeAsO$_{0.83}$F$_{0.17}$. Straight lines are guides to the eye.

Figure 4. Hydrostatic pressure dependence of the relative lattice parameters c/a of SmFeAsO$_{1-x}$F$_x$ for x=0.0 and 0.17. Straight lines are guides to the eye.

Figure 5. Hydrostatic pressure dependence of Sm phonon frequencies of SmFe$_{0.92}$Co$_{0.08}$AsO. Straight lines are guides to the eye.

Figure 6. Hydrostatic pressure dependence of the Sm phonon width of SmFe$_{0.92}$Co$_{0.08}$AsO. Straight lines are guides to the eye.

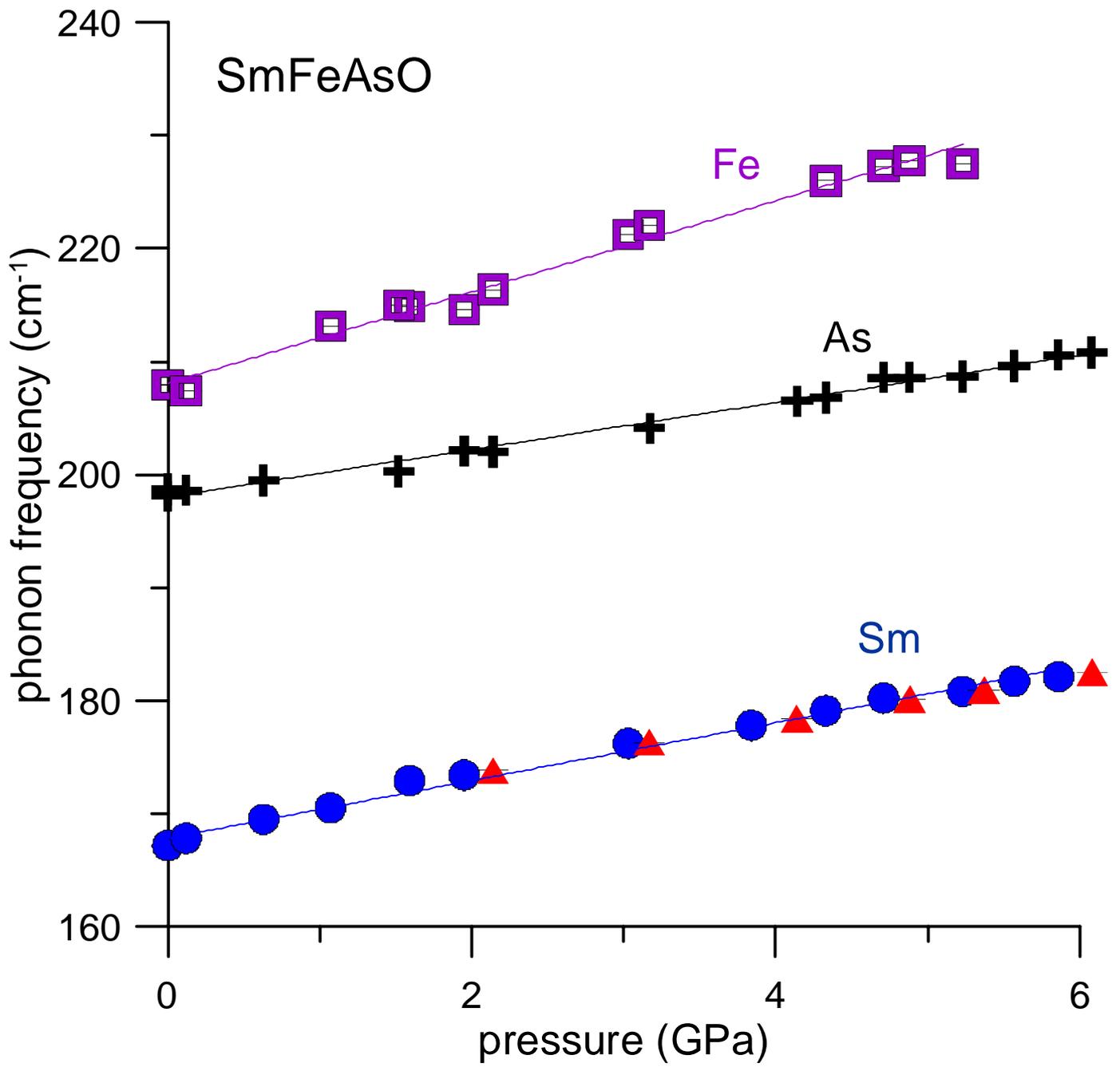

Fig.1

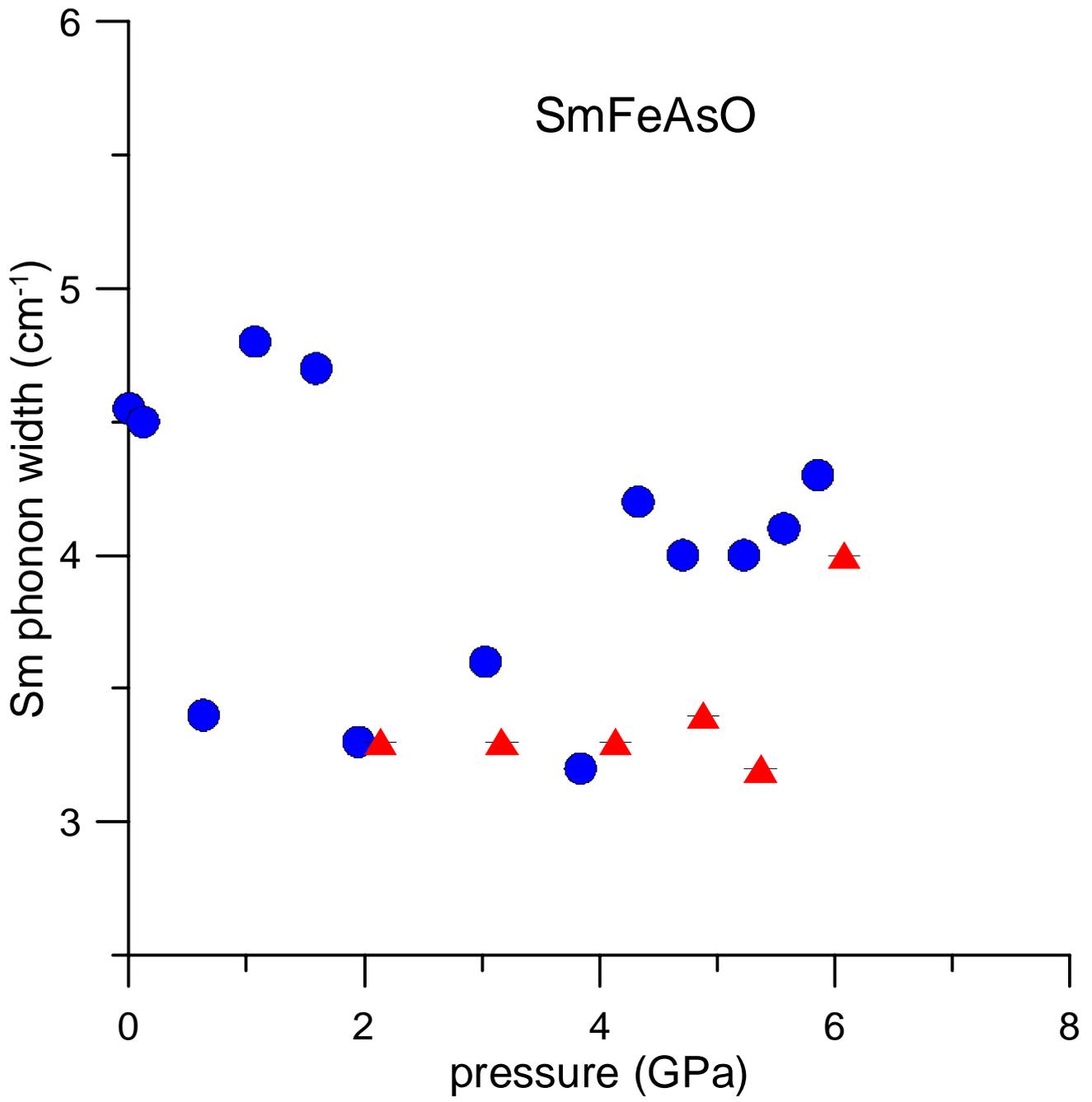

Fig.2

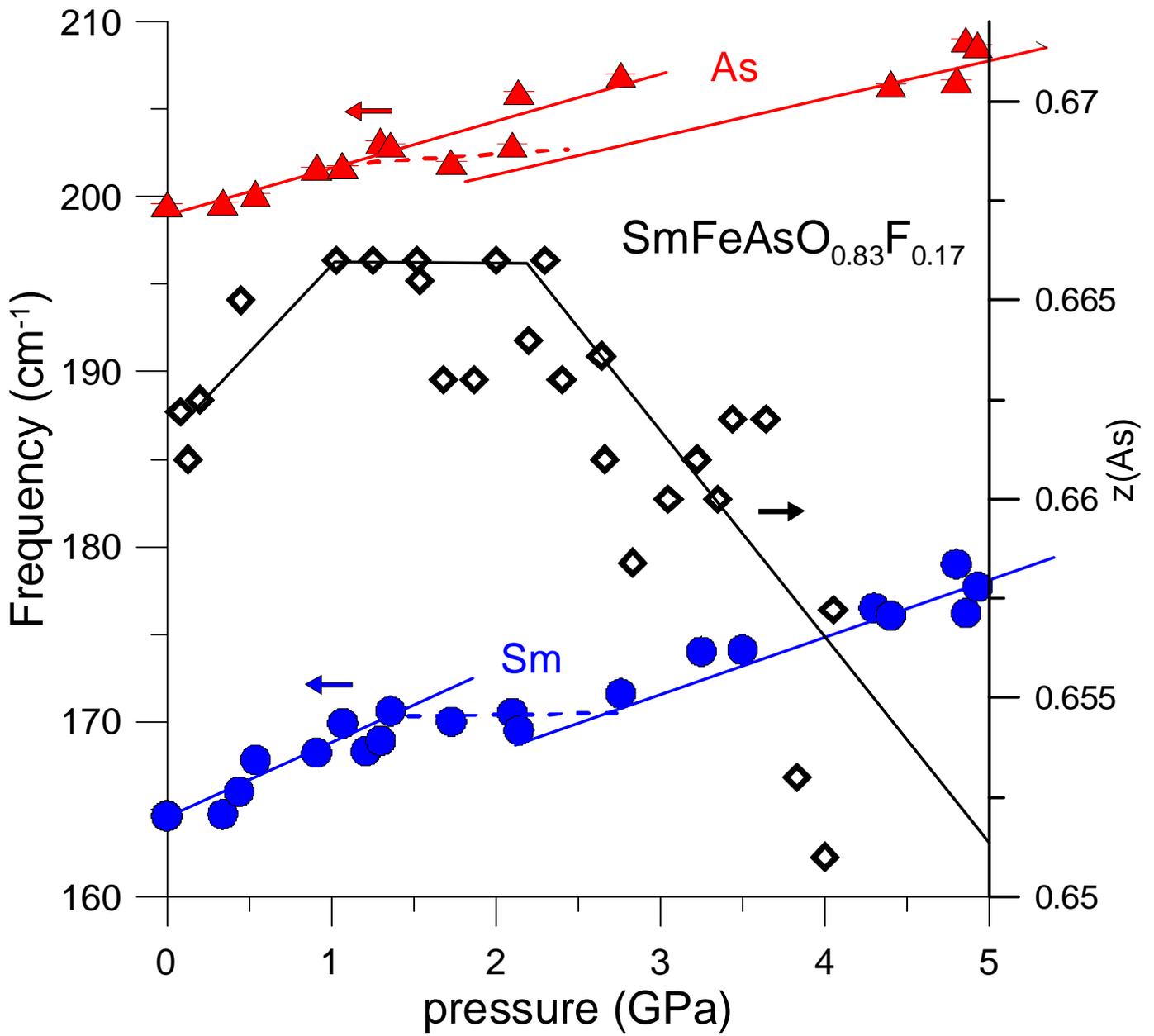

Fig.3

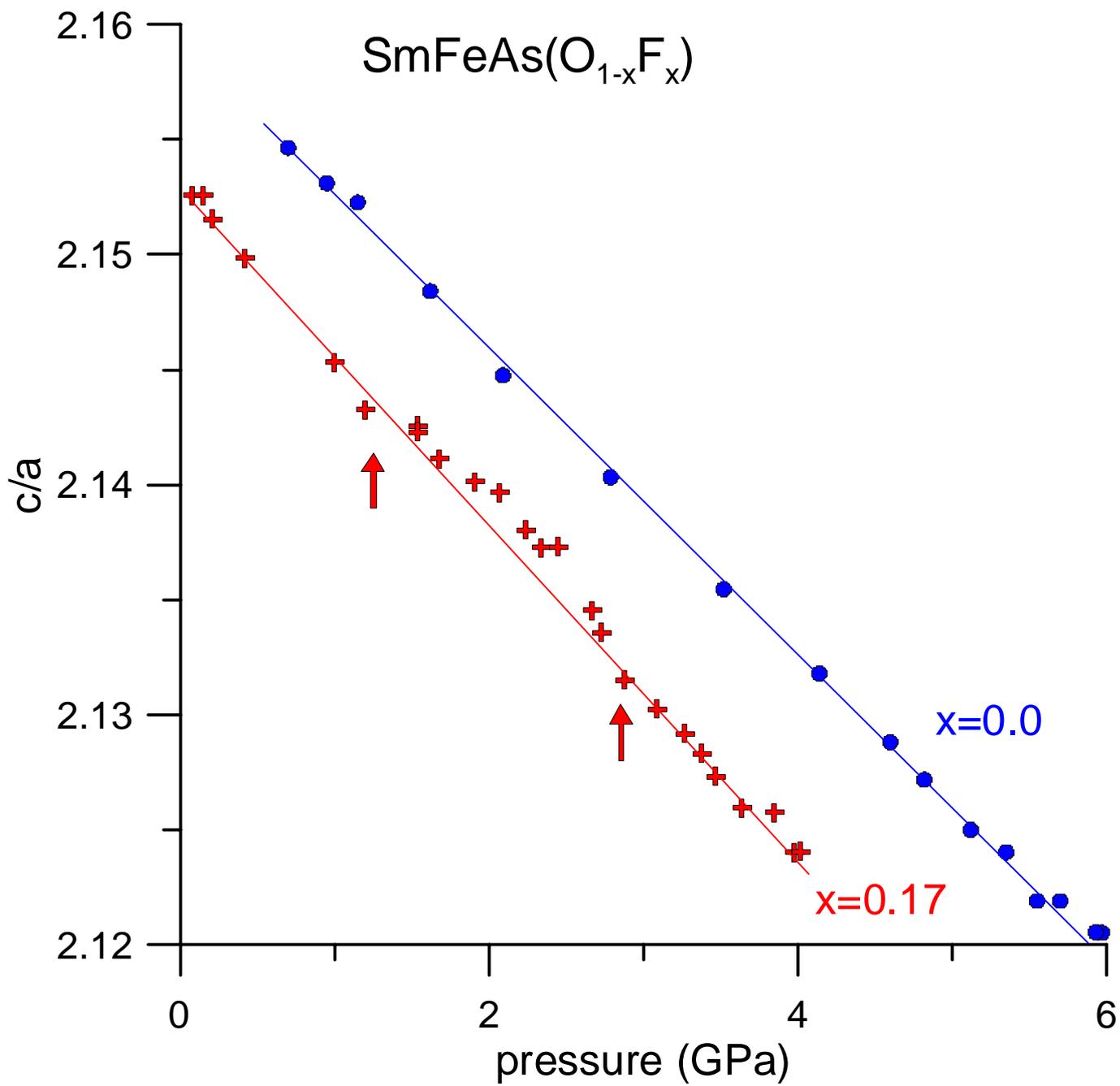

Fig.4

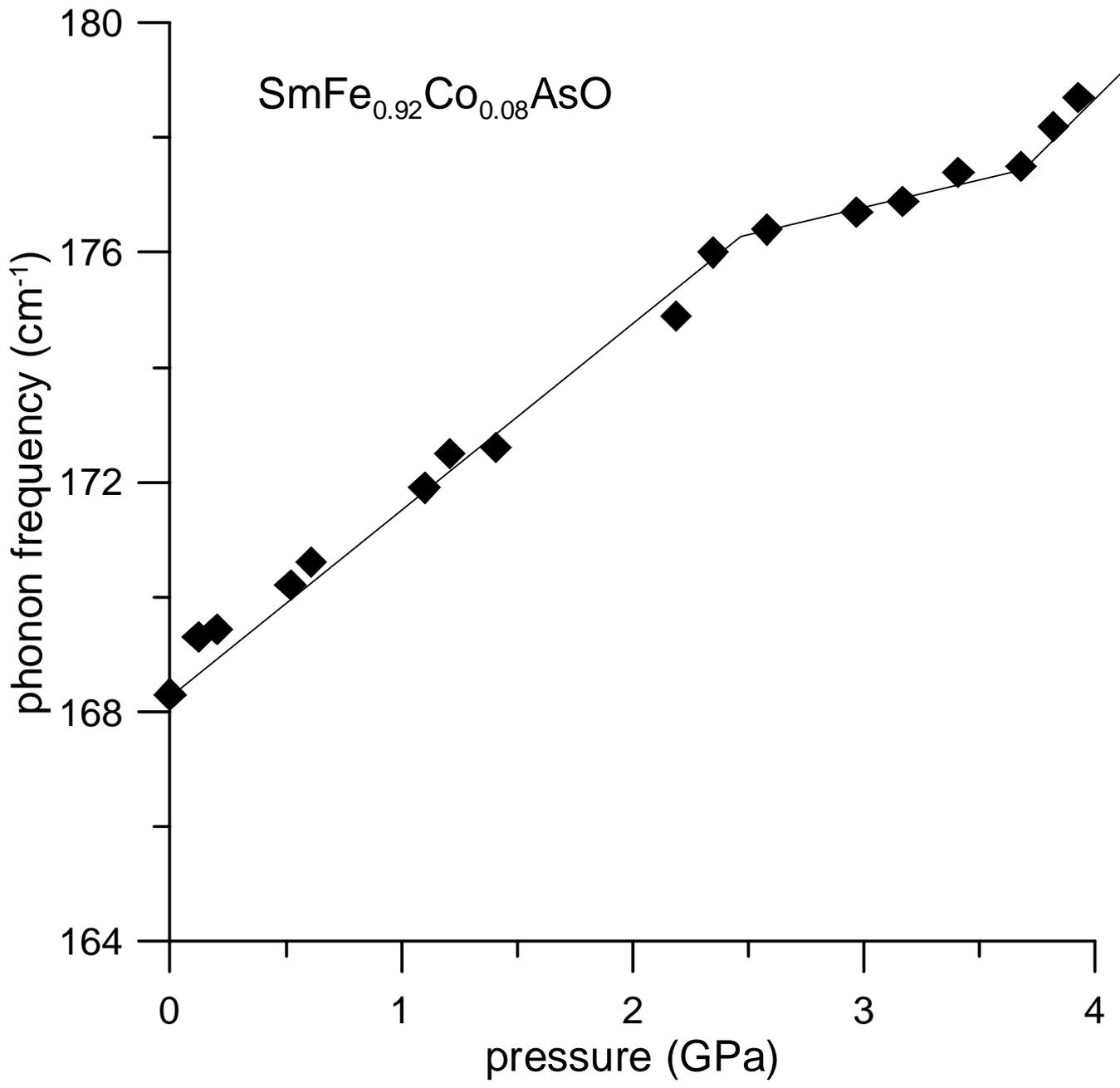

Fig.5

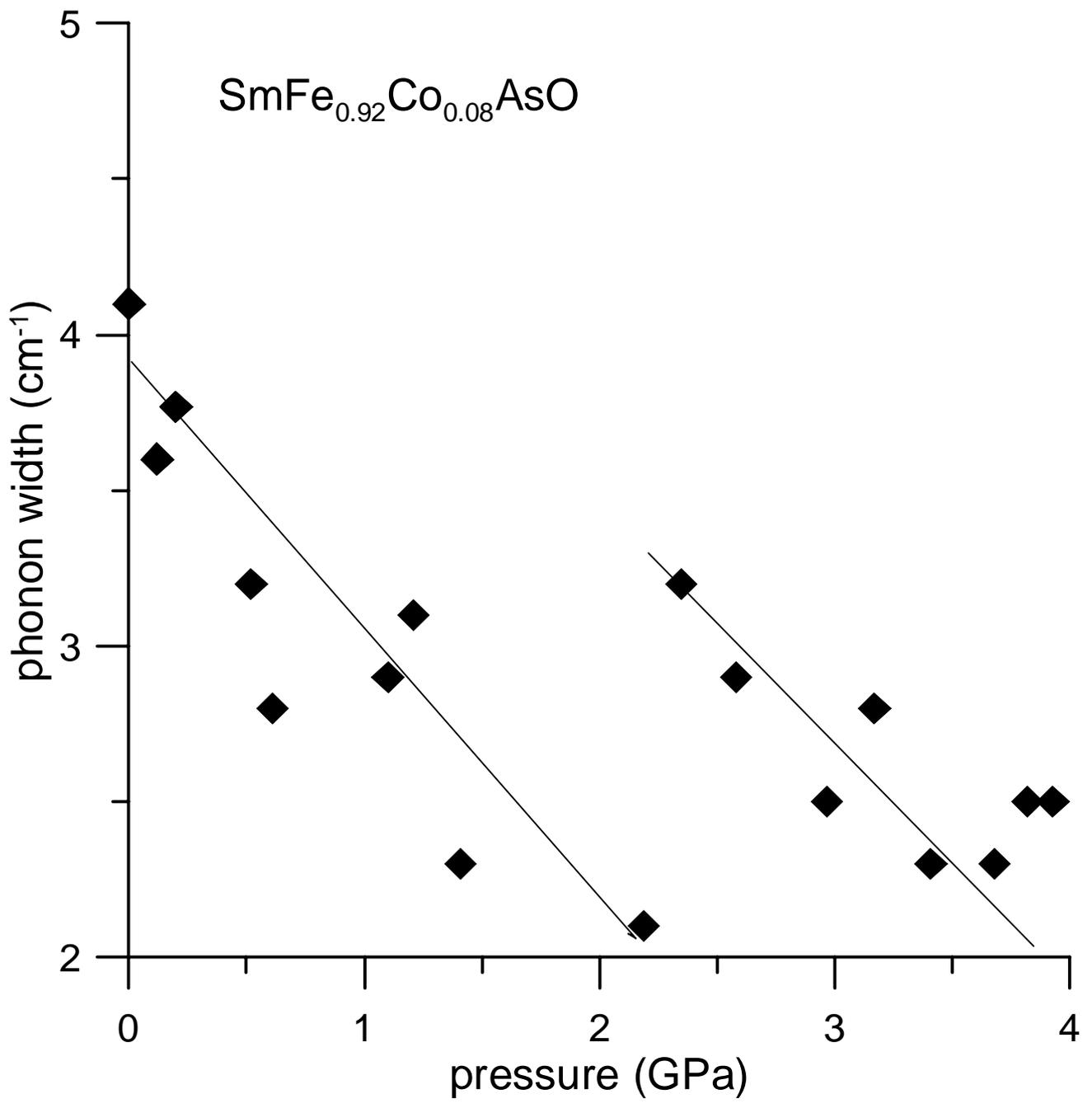

Fig.6